\documentclass[11pt,a4paper]{article}
\usepackage{jcappub}

\usepackage[latin1]{inputenc}
\usepackage[english]{babel}
\usepackage{amsfonts}
\usepackage{amsmath}    
\usepackage{amssymb}
\usepackage{amsthm}
\usepackage{slashed}
\usepackage[figure,table]{hypcap} 
\hypersetup{
   bookmarksnumbered,
   pdfstartview={FitV},
   pdfpagemode={UseOutlines},
   pdfauthor={Torsten Bringmann, Jasper Hasenkamp, J\"orn Kersten},
   pdftitle={},
   pdfsubject={scientific publication}
   ,citecolor={blue},
   linkcolor={blue},
   urlcolor={blue},
   filecolor={blue}
} 

\newcommand{\be}{\begin{equation}}
\newcommand{\ee}{\end{equation}}
\newcommand{\Uoney}{\ensuremath{{U}(1)_{{Y}}}}
\newcommand{\Uonex}{\ensuremath{{U}(1)_{{X}}}}
\newcommand{\Neff}{N_\text{eff}}
\newcommand{\order}[1]{\ensuremath{\mathcal{O}\left(#1\right)}}
\newcommand{\mplanck}{\ensuremath{M_{\text{pl}}}}
\newcommand{\fb}[3]{\left(\frac{#1}{#2}\right)^{#3}}

\begin{document}
 
\date{February 24, 2014}

\title{Tight bonds between sterile neutrinos and dark matter}

\author[a]{Torsten Bringmann}
\emailAdd{Torsten.Bringmann@fys.uio.no}
\affiliation[a]{Department of Physics, University of Oslo, Box 1048, NO-0316 Oslo, Norway}

\author[b]{Jasper Hasenkamp}
\emailAdd{Jasper.Hasenkamp@nyu.edu}
\affiliation[b]{\mbox{CCPP, Department of Physics, New York University, New York, NY 10003, USA}}

\author[c]{J\"{o}rn Kersten}
\emailAdd{Joern.Kersten@desy.de}
\affiliation[c]{{II.} Institute for Theoretical Physics, University of Hamburg, 
Luruper Chaussee 149, DE-22761 Hamburg, Germany}

\arxivnumber{1312.4947}

 \abstract{
Despite the astonishing success of standard $\Lambda$CDM cosmology, there is mounting 
evidence for a tension with observations at small and intermediate scales. We introduce a 
simple model where both cold dark matter (DM) and sterile neutrinos are charged under a new 
\Uonex\ gauge interaction.  The resulting DM self-interactions  resolve the tension with the 
observed abundances and internal density structures of dwarf galaxies. At the same time, the 
sterile neutrinos can account for both the small hot DM component favored by cosmological 
observations and the neutrino anomalies found in short-baseline experiments.
}

\maketitle

\section{Introduction}

The nature of the dark matter (DM) in the Universe is one of the outstanding puzzles of science.
Intensive research over the past decades led to the current standard DM paradigm of cold, 
collisionless particle DM (CDM). Alongside the remarkable success of the standard cosmological 
model, known as $\Lambda$CDM, in describing the Universe at large scales, a number of 
problems became increasingly solid over the past years driven by the increase in precision of 
cosmological and astrophysical observations. 

On small cosmological scales those are most prominently: 1) \textit{Missing satellites} -- the DM 
halo of the Milky Way (MW) should contain many more dwarf-sized subhalos (satellites) than 
observed~\cite{Klypin:1999uc,Kravtsov:2009gi}. Beyond the MW, observed galaxy 
luminosity and H{\sc i}-mass functions show shallower faint-end slopes than predicted
\cite{Zavala:2009ms}. 2) \textit{Cusp vs.~core} -- low surface brightness  and dwarf 
galaxies seem to have cored inner density profiles,  at odds with CDM cusps predicted by 
simulations~\cite{deNaray:2011hy,Walker:2011zu}. 3) \textit{Too big to fail} -- the observed 
brightest satellites of the MW attain their maximum circular velocity at too large radii in comparison 
to the densest and most massive satellites found in simulations, i.e.~the latter have no counterpart 
in observations even though they should be very efficient in forming 
stars~\cite{BoylanKolchin:2011de,BoylanKolchin:2011dk}.
Both astrophysical  \cite{Efstathiou:1992sy,Pfrommer:2011bg,Silk:2010aw,Shapiro:2003gxa,MacLow:1998wv,Hopkins:2011rj,Pontzen:2011ty,VeraCiro:2012na,Wang:2012sv} and 
DM-related \cite{Spergel:1999mh,Bode:2000gq,Dalcanton:2000hn,Zentner:2003yd,Sigurdson:2003vy,Kaplinghat:2005sy,Borzumati:2008zz,Boehm:2000gq,Kaplinghat:2000vt,Hu:2000ke,Lovell:2011rd,Feng:2009hw,Buckley:2009in,Loeb:2010gj,Vogelsberger:2012ku,Tulin:2013teo} 
solutions are actively discussed, 
but the first {\it simultaneous} solution has only been advanced rather recently \cite{Aarssen:2012fx} (see also Ref.~\cite{Marsh:2013ywa} for a two-component DM model that augurs well in this respect).

On intermediate scales the observed galaxy cluster mass function and galaxy shear power 
spectra,  both tracing the matter power spectrum, are inconsistent with observations of the 
cosmological microwave background (CMB). Also the current expansion 
rate of the Universe $H_0$ appears inconsistent with CMB data. More generally, 
``local'' measurements of cosmological parameters, probing cosmology at ``low'' redshifts 
$z \lesssim 10$, are inconsistent with CMB observations that are particularly sensitive  to high 
redshifts $z\gtrsim 1000$~\cite{Ade:2013zuv}. A mixed DM model, which adds a small hot dark 
matter (HDM) component  to the dominant CDM,
resolves those inconsistencies and thus reconciles the low and high redshift universe 
\cite{Wyman:2013lza,Hamann:2013iba,Battye:2013xqa,Gariazzo:2013gua}. This requires 
an effective number of additional neutrino species $\Delta\Neff$ and an effective HDM mass of   
\cite{Hamann:2013iba} 
\be
\label{HDMsignal}
\Delta\Neff|_\text{cmb} = 0.61 \pm 0.30\,,\; m_\text{hdm}^\text{eff}
=(0.41 \pm 0.13) \,\text{eV}\,.
\ee

Here, we extend the standard model of particle physics (SM) by a spontaneously broken 
\Uonex\ gauge theory and 
introduce a sterile neutrino that is not charged under SM interactions.
We demonstrate  that all aforementioned problems of standard cosmology are resolved by 
coupling both CDM and sterile neutrinos, the latter automatically being promoted to the desired 
HDM component, to a \Uonex\ gauge boson of \order{\text{MeV}} mass. 
This can be achieved for parameters that resolve the 
anomalies~\cite{Aguilar:2001ty,Aguilar-Arevalo:2013pmq,Mention:2011rk,Huber:2011wv,Hampel:1997fc,Abdurashitov:2005tb,Giunti:2010zu} 
reported in short-baseline neutrino oscillation experiments,
in particular a sterile neutrino mass of $\sim 1$\,eV.

This article is organized as follows. We first set up our model  
and give a brief overview of the cosmological evolution of the non-standard degrees 
of freedom introduced here, as sketched in Fig.~\ref{fig:overview}. We continue by showing that the 
properties of the thermally produced CDM sector can provide a solution  to all $\Lambda$CDM 
problems at small scales. Next, we demonstrate that our sterile neutrino HDM component can {\it 
simultaneously} satisfy the cosmologically favored values  stated above and describe the 
anomalies in short-baseline oscillation experiments. We conclude with a discussion and 
an outlook.

\section{Model setup}

We consider the extension of the SM gauge group, 
${G}_\text{SM} = SU(3)_c \times SU(2)_{L} \times \Uoney$, by an Abelian gauge 
symmetry \Uonex\ with corresponding gauge boson $V$. 
We introduce a Dirac fermion $\chi$ at the TeV scale, which will form the CDM, and two right-
handed neutrinos $\nu_{R_{1,2}}$. Those new particles are neutral under $G_\text{SM}$ but carry 
\Uonex\ charges, while the SM particles are neutral under  \Uonex. Anomaly cancellation requires 
the $\nu_{R_{1,2}}$ to carry charges of opposite sign with equal absolute value; for concreteness, 
we take the charges of $(\chi,\nu_{R_{1}},\nu_{R_{2}})$ to be $(1,X_{\nu_R},-X_{\nu_R})$.

We further assume that the \Uonex\ is spontaneously broken at the MeV scale by the vacuum 
expectation value (VEV) $v_\Theta$ of a complex Higgs field $\Theta$, which is a representation 
$(1,0,2 X_{\nu_R})$ under $SU(2)_L \times \Uoney \times \Uonex$, while the Higgs field 
$\phi$ responsible for the electroweak symmetry breaking 
is a $(2,1/2,0)$.
Another complex scalar $\xi$, with charges $(1,0,X_{\nu_R})$ and VEV $v_\xi < v_\Theta$, is introduced 
to enable active-sterile neutrino mixing.

After symmetry breaking the low-energy, effective Lagrangian of our theory reads
\begin{equation}
\label{L}
 \mathcal{L}= \mathcal{L}_\text{SM} + \mathcal{L}_{R} + \mathcal{L}_x  + 
  \mathcal{L}_\text{kin.\@ mix.} +\mathcal{L}_\text{Higgs}\, .
\end{equation}
Here, $\mathcal{L}_\text{SM}$ denotes SM terms and $\mathcal{L}_{R}$ contains 
\begin{align}
 \label{Lnumass}
\mathcal{L}_{R}  \supset& -\frac{1}{2} \overline{\nu^c_{R}}_{{ }_1} M_1 \nu_{R_1}  
- \frac{1}{2} \overline{\nu^c_{R}}_{{ }_2} M_2 \nu_{R_2}
- \overline{\nu_{R}^c}_{{ }_1} M_{RR} \nu_{R_2} -  \overline{\nu_L} M_{LR} \nu_{R_1} + \text{h.c.} \,,
\end{align}
in addition to kinetic terms and Majorana mass terms for the SM neutrinos.
The active-sterile neutrino mixing arises from a dimension-5 operator with 
$M_{LR} \sim v_\phi v_\xi/\Lambda$, suppressed by a scale $\Lambda$ defined by the UV completion of  
the theory. The mass eigenstates $(\nu_1, \nu_2, \nu_3, N_1, N_2)$ are mixtures of the flavor 
eigenstates $(\nu_e, \nu_\mu, \nu_\tau, \nu_{R_1}^c, \nu_{R_2}^c)$. 
With the short-baseline anomalies in mind, Yukawa couplings between $\nu_{R_{1,2}}$ and 
$\Theta$ are chosen such that 
$m_{N_2}\!\sim\! M_2\!\sim\!\text{MeV}\gg  m_{N_1}\!\sim\!M_1 \!\sim\!\text{eV}$. 
We note that even a very
small mixing between 
$\nu_{R_{1}}$ and $\nu_{R_{2}}$, with $M_{RR}/M_2\!\gtrsim\!10^{-6}$, allows the 
cosmologically fast decay of $\nu_{R_2}$. 

Terms in $\mathcal{L}$ related to  $V$ and the new fermions are
\begin{align}
\label{Lx}
 \mathcal{L}_x ={}& \bar \chi (i \slashed \partial - m_\chi) \chi 
- \frac{1}{4} F_{\mu\nu}^x F^{x\mu\nu} - \frac{1}{2} m_V^2 V_\mu V^\mu \\ 
&-g_X V_\mu\left(X_{\nu_R}\overline{ \nu_{R}}_{{ }_1} \gamma^\mu  \nu_{R_1} 
-  X_{\nu_R}\overline{ \nu_{R}}_{{ }_2} \gamma^\mu  \nu_{R_2} 
+ \bar \chi \gamma^\mu \chi \right) , \nonumber
\end{align}
where $g_X$ denotes the \Uonex\ gauge coupling. To ensure the stability of $\chi$ we might 
impose a discrete $Z_2$ symmetry under which only $\chi$ is assigned a negative parity.
The symmetries also allow a kinetic mixing term
 $\mathcal{L}_\text{kin.\ mix.} = -\frac{\epsilon}{2} F_{\mu\nu}^x F^{\mu\nu}$, where 
 $F^{x\mu\nu}$ ($F^{\mu\nu}$) denotes the \Uonex\  (electromagnetic) field strength tensor. We 
 assume $\epsilon\ll1$ to satisfy the severe existing constraints on this parameter \cite{Jaeckel:2013ija,Kaplinghat:2013yxa}.

We refer to Ref.~\cite{Ahlers:2008qc} for a general discussion of the Higgs sector for  $\Theta$ 
and $\phi$ as contained in $\mathcal{L}_\text{Higgs}$, adopting that $m_V$ and the mass 
of the new light  Higgs boson $h_x$ are of the same order of magnitude in the relevant cases. 
The ``Higgs portal'' term in 
$\mathcal{L}_\text{Higgs} \supset \kappa |\phi|^2|\Theta|^2 \supset \frac{\kappa}{4} v_\phi \phi \Theta^2 \simeq \frac{\kappa}{4} v_\phi h h_x^2$, 
where we have assumed a negligible mixing between $h_x$ and the SM-like Higgs $h$ in 
the last step, 
connects the SM and the new \Uonex\ sector.
For simplicity, we assume the couplings of the additional portal terms $|\phi|^2 |\xi|^2$ and $|\Theta|^2 |\xi|^2$ to be negligibly small.

\section{Thermalization via the Higgs portal and decoupling of the Dark Sector}

\begin{figure}[t!]
 \label{fig:overview}
\centering
 \includegraphics[width= 0.5\columnwidth]{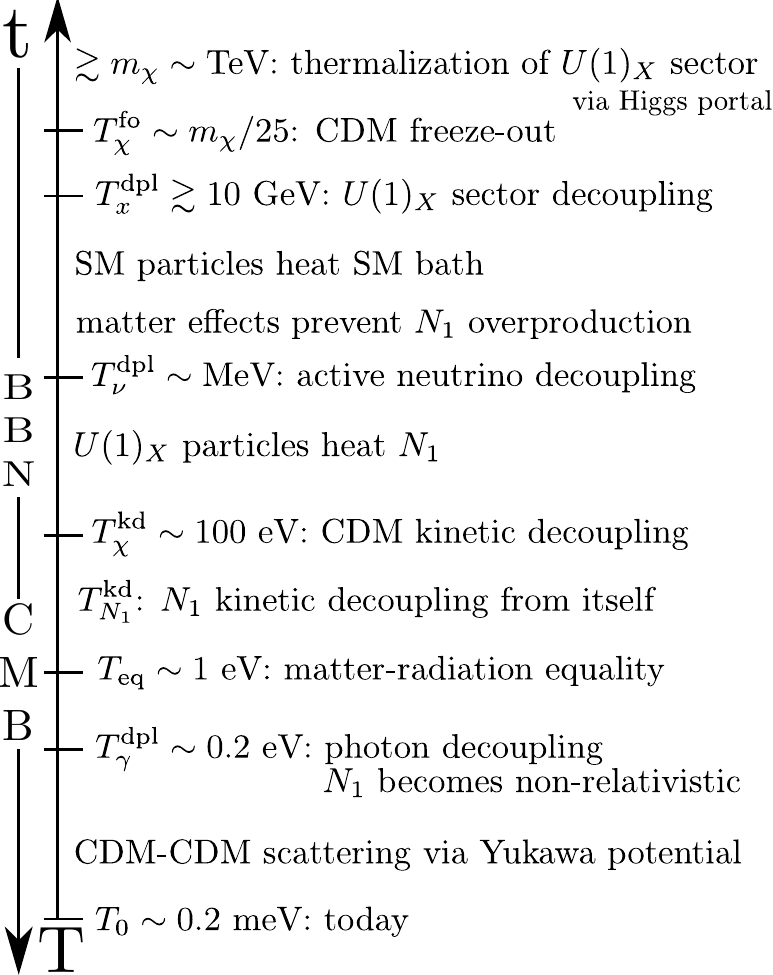}
\caption{Schematic overview of the cosmology implied by the model defined in Eq.~\eqref{L}. }
 \end{figure}

In Fig.~\ref{fig:overview} we provide  a schematic overview of the cosmology arising from our 
model represented by Eq.~\eqref{L}. Before electroweak symmetry breaking,  the 4-scalar 
interaction $\kappa |\Theta|^2 |\phi|^2$ keeps the \Uonex\ sector in thermal equilibrium with the SM 
bath if the thermalization rate 
$\Gamma_\text{th}= n_{h_x} \langle \sigma_\text{th} v_\text{rel} \rangle \sim 10^{-3}\kappa^2 T$ is larger than the 
expansion rate $H\sim 10\, T^2/\mplanck$.  In those expressions, $n_{h_x}$ denotes the number 
density of $h_x$ and $\langle \sigma_\text{th} v_\text{rel} \rangle$ the thermally averaged annihilation 
cross section of $h_x$ pairs. 
If we, e.g., require thermal equilibrium at temperatures below $10\,\text{TeV}$, i.e.~above the CDM mass $m_\chi$, we obtain 
a lower bound on the Higgs portal coupling of $\kappa \gtrsim 10^{-6}$. 
After electroweak symmetry breaking the relevant process becomes 
$h_x  h_x\stackrel{h}{\rightarrow} f  \bar f$, controlled by the $\frac{\kappa}{4} v_\phi h h_x^2$ coupling.
 The thermalization rate is then 
$\Gamma_\text{th} \sim 10^{-3}\,\kappa^2 T^3 m_f^2/m_h^4$, with $f$ corresponding to the heaviest 
relativistic SM fermion, so  the decoupling temperature becomes 
$T_x^\text{dpl} \sim 10^3 m_h^4/(\kappa^2 m_f^2 \mplanck)$.
For details on thermalization via the Higgs portal we refer to
\cite{McDonald:1993ex,Burgess:2000yq,Barger:2007im}, where thorough calculations of $h_x$ 
abundances for $m_{h_x} \sim \text{TeV}$ were performed (while in our 
case $h_x$ decouples relativistically).

The particles in the dark sector are  tightly coupled to each other due to the \Uonex\ 
interaction, and more weakly to the SM via the Higgs portal. Once the latter ceases to be effective, 
the whole \Uonex\ sector therefore decouples from the SM bath and  entropy is  
conserved separately in the two sectors. Whenever a particle in equilibrium 
becomes non-relativistic it thus heats its bath, thereby increasing the temperature by a factor  
$(g_{\ast,\nu/N_1}^\text{before}/g_{\ast,\nu/N_1}^\text{after})^\frac{1}{3}$,
where $g_{\ast,i}$ counts the effective degrees of freedom (d.o.f.) determining the entropy 
density of the  sector in thermal equilibrium with the species $i$.
The  non-standard contribution to the radiation density is then given by
\be
  \Delta N_{\rm eff}(T)
  = \frac{T_{N_1}^4}{T_\nu^4}=
  \left.\left(\frac{g_{*,\nu}}{g_{*,N_1}}\right)^\frac43\right|_T\left.\left(\frac{g_{*,N_1}}{g_{*,\nu}}\right)^\frac43\right|_{T^{\rm dpl}_x}.
\ee
The {\it maximal} possible value of this quantity at the onset of big bang nucleosynthesis (BBN), at 
$T\sim1$\,MeV, is then obtained if {\it all} new particles but the light sterile neutrino, $N_1$, have 
become non-relativistic by then. This results in 
\be
\label{DNeffbbn}
\Delta\Neff|_\text{bbn}^\text{max} \simeq \left[58.4/g_{*,\nu}(T^{\rm dpl}_x)\right]^\frac43 ,
\ee
well within  bounds from BBN \cite{Mangano:2011ar,Izotov:2013waa,Aver:2013wba} for $T^{\rm dpl}_x\gtrsim1$\,GeV.

\section{Self-interacting CDM}
 
 \begin{figure}[t!]
 \label{fig:mv_mchi}
 \centering
  \includegraphics[width= 0.9\columnwidth]{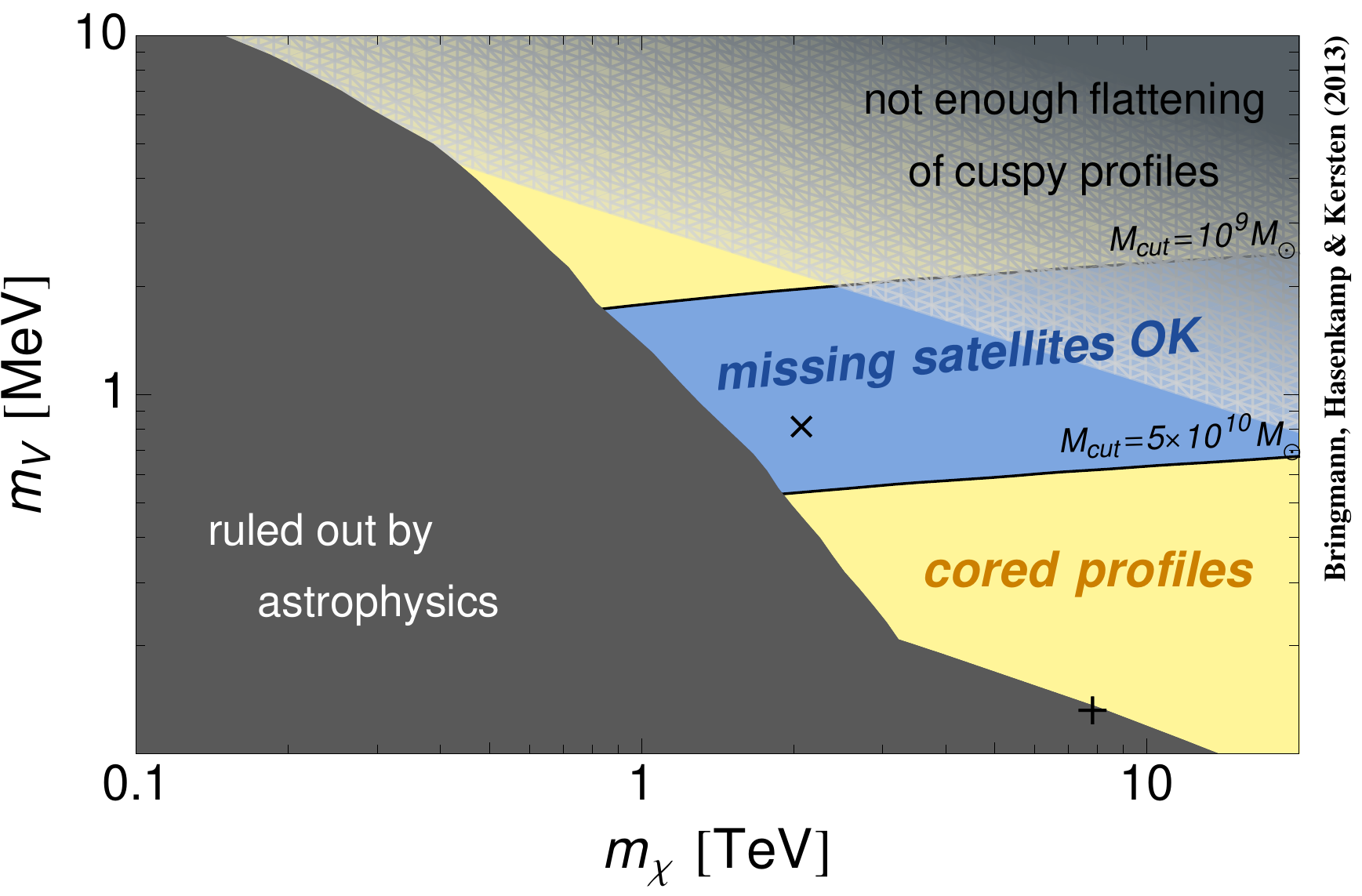}
 \caption{In the yellow area, the CDM self-interaction is strong enough to flatten density {\it cusps} 
 in the inner parts of (dwarf) galaxies \cite{Loeb:2010gj} and likely also solves  the 
 {\it too big to fail} problem (as explicitly demonstrated in $N$-body simulations for parameter 
 values corresponding to the crosses \cite{Vogelsberger:2012ku}).
  The dark area is excluded by astrophysics  \cite{Gnedin:2000ea,Balberg:2002ue,Buckley:2009in,Loeb:2010gj}.
 The blue band addresses the {\it missing satellites} problem \cite{Aarssen:2012fx}, with a normalization that -- according 
 to Eq.~(\ref{tkd}) -- is proportional to  $m_V\propto X_{\nu_R}^{1/2}(T_{N_1}/T)^{3/2}_{\rm kd}$. Here, we show for 
 reference the case of $X_{\nu_R}=0.2$ and $(T_{N_1}/T)_{\rm kd}^4=0.46$. }
 \end{figure}

At high temperatures, the DM particles are kept in chemical equilibrium via 
$\chi\chi\leftrightarrow VV$ (for unit sterile neutrino charges, $X_{\nu_R}\sim1$, also the annihilation into $\overline{\nu_R}{\nu_R}$, $h_xh_x$ and $\xi^*\xi$ via a virtual $V$ becomes important). For TeV-scale DM the number density freezes out at sufficiently early 
times ($T_\chi^\text{fo} \sim m_\chi/25$) to still have $T_V=T$.  Assuming for simplicity $X_{\nu_R}\ll1$, the  CDM relic 
density then becomes
\begin{equation}
 \label{Ochi}
 \Omega_\text{cdm} h^2 = 2\, \Omega_\chi h^2  \sim 0.11
\fb{0.67}{g_X}{4}
\fb{m_\chi}{\text{TeV}}{2} 
\end{equation}
up to \order{1} corrections due to  
the Sommerfeld effect \cite{ANDP:ANDP19314030302}, which we fully take into account \cite{Aarssen:2012fx}.
This fixes $g_X$ for a given $m_\chi$ throughout this work.

Kinetic decoupling \cite{Bringmann:2009vf} of $\chi$ happens much later and is determined by the elastic 
scattering rate for $\chi N_1\leftrightarrow \chi N_1$.  For a thermal distribution of sterile neutrinos, 
the decoupling temperature is given by \cite{Aarssen:2012fx}
\be
 \label{tkd}
 T_\chi^{\rm kd} \simeq \frac{62\,{\rm eV}}{X_{\nu_R}^\frac12 g_X}
  \left(\frac{T}{T_{N_1}}\right)^\frac32_{\rm kd}\left(\frac{m_\chi}{\rm TeV}\right)^\frac14
  \left(\frac{m_V}{\rm MeV}\right) ,
\ee
which translates into a cutoff in the power spectrum of matter density perturbations at 
$M_{\rm cut}\sim  1.7\times10^8(T_\chi^{\rm kd}/{\rm keV})^{-3}M_\odot$. We note that the light 
mass eigenstates $\nu_i$ also acquire a \Uonex\ charge from their $\nu_R^c$ component; this will 
further  lower $T_\chi^{\rm kd}$ if $\sin\theta\gtrsim (T_{N_1}/T_\nu)^{3/2}$.

After structure formation, the \Uonex-induced Yukawa potential  produces galaxy cores that match 
the observed velocity profiles of massive MW satellites, solving 
\textit{cusp vs.\@ core} \cite{Feng:2009hw,Loeb:2010gj} and \textit{too big to fail} \cite{Vogelsberger:2012ku}, while avoiding constraints on 
DM self-interactions on larger scales \cite{Loeb:2010gj}. At the same time, the late kinetic 
decoupling addresses the  \textit{missing satellites} by suppressing the matter power spectrum at 
dwarf galaxy scales~\cite{Aarssen:2012fx} (see also Refs.~\cite{Boehm:2000gq,Chen:2001jz,Boehm:2004th,Shoemaker:2013tda}). In Fig.~\ref{fig:mv_mchi}, we show the desired 
parameter space for $m_V$ and $m_\chi$ (based on Ref.~\cite{Aarssen:2012fx}, but using an  
improved parameterization \cite{upcoming} of the Yukawa scattering cross section \cite{khrapak1,khrapak2,khrapak3,Feng:2009hw,Tulin:2012wi}).
The blue band, in particular, shows the range of masses that allow a solution of
the missing satellites problem. Note that its normalization depends on the choice of
$X_{\nu_R}$ and $(T_{N_1}/T)_{\rm kd}$, whereas the $m_\chi$-dependence is uniquely determined by Eq.~(\ref{tkd}) and 
 the form of $g_X(m_\chi)$ that corresponds to the correct thermal CDM relic density, cf.~Eq.~(\ref{Ochi}).
The dark area is excluded by the requirements to not disrupt galactic satellites and to avoid a gravothermal catastrophe  \cite{Gnedin:2000ea,Balberg:2002ue,Buckley:2009in,Loeb:2010gj}.

\section{The HDM component}

 \begin{figure}[t!]
 \label{fig:hdm}
 \centering
  \includegraphics[width= 0.9\columnwidth]{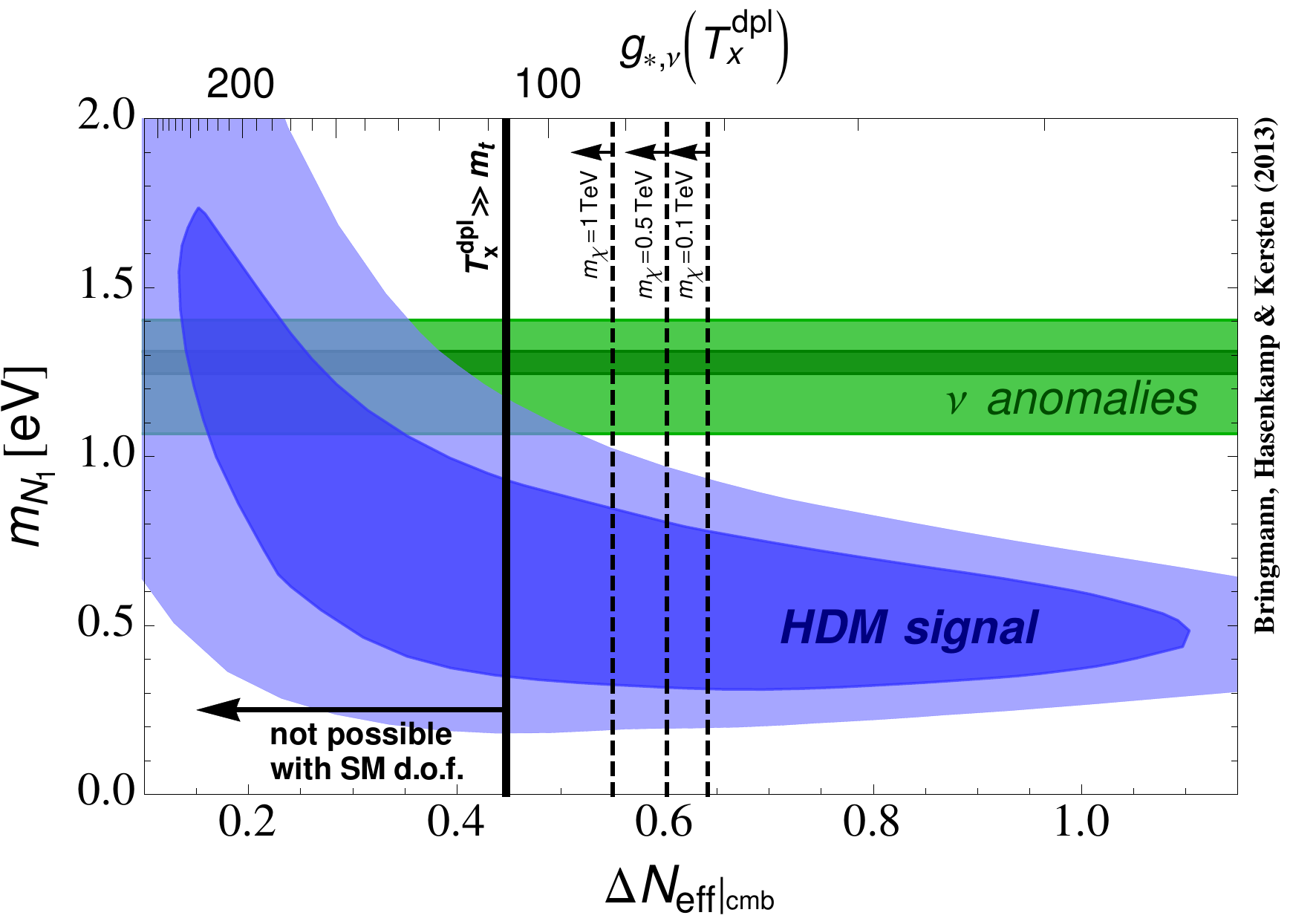}
 \caption{Sterile neutrino mass $m_{N_1}$ vs.~late-time additional relativistic 
 d.o.f.~$\Delta\Neff|_\text{cmb}$ and SM d.o.f.~at decoupling of the \Uonex\ sector, 
 cf.~Eq.~(\ref{DNeffbbn}). Shaded areas correspond, at $1\sigma$ and  $2\sigma$ 
 respectively, to the HDM  signal~\cite{Hamann:2013iba} and values of $m_{N_1}$ favored by 
 the neutrino anomalies~\cite{Giunti:2013aea,Giunti:2013waa}. Dashed lines indicate the 
 minimal value of $\Delta\Neff|_\text{cmb}$ compatible with a CDM mass of, from right to left, 
 $m_\chi=100,500,1000$\,GeV\@. Parameter values to the left of the solid line are not achievable 
 in the minimal scenario studied here.}
 \end{figure}

We will now address the question whether the $N_1$ population in our model can account for the 
cosmologically preferred HDM component  
\cite{Wyman:2013lza,Hamann:2013iba,Battye:2013xqa,Gariazzo:2013gua}.
In the absence of any significant additional $N_1$ production mechanism, see the discussion 
further down, we simply have 
\be
\label{dNsimp}
\Delta\Neff|_\text{cmb} = \Delta\Neff|_\text{bbn}^\text{max}\,.
\ee 
From the 
definition $m_\text{hdm}^\text{eff} \equiv\left[{T_{N_1}}/T_\nu^{\Lambda{\rm CDM}}\right]^3 m_{N_1}={11}/{4}\left[{T_{N_1}}/T\right]^3 m_{N_1}$  
it furthermore follows that 
\be
\label{mhdmsimp}
m_\text{hdm}^\text{eff} = (\Delta\Neff|_\text{cmb})^\frac34 m_{N_1}\,.
\ee 
By choosing the right decoupling temperature in Eq.~\eqref{DNeffbbn}, which in our model 
corresponds to  adjusting $\kappa$, we can then reproduce  Eq.~\eqref{HDMsignal}. 
%
This is demonstrated in Fig.~\ref{fig:hdm} where we show the allowed region of  
$\Delta\Neff|_\text{cmb}$ and $m_\text{hdm}^\text{eff}$ \cite{Hamann:2013iba} in terms of 
$g_{*,\nu}(T^{\rm dpl}_x)$ and $m_{N_1}$. 

The thermal production of the CDM component
as treated here requires $T^{\rm dpl}_x \lesssim T_\chi^\text{fo}\sim m_\chi/25$; 
in Fig.~\ref{fig:hdm}, this corresponds to the area to the right of the dashed lines 
 for various values of $m_\chi$.
On the other hand, $g_{*,\nu}(T^{\rm dpl}_x)$ 
cannot exceed the full number of SM d.o.f.~even for very early \Uonex\ decoupling. Taken together,
this points to $0.2\,\text{eV}\lesssim m_{N_1}\lesssim1.2\,\text{eV}$.

\section{Neutrino anomalies}

Oscillation experiments observing neutrinos from accelerators
\cite{Aguilar:2001ty,Aguilar-Arevalo:2013pmq},
reactors~\cite{Mention:2011rk,Huber:2011wv} (but see~\cite{Hayes:2013wra}), and radioactive sources
\cite{Hampel:1997fc,Abdurashitov:2005tb,Giunti:2010zu}
reported anomalies that may
indicate the existence of sterile neutrinos with a mass
squared difference $\Delta m^2 \sim 1\,\text{eV}^2$ to the SM neutrinos.
In Fig.~\ref{fig:hdm}, we show the $1\sigma$ and $2\sigma$ ranges for $\Delta m^2$
from \cite{Giunti:2013aea,Giunti:2013waa}
for orientation, assuming $m_{N_1}^2=\Delta m^2$.  These ranges were obtained from a global fit of
oscillation data assuming the existence of a single sterile neutrino 
(note that it is being debated to what extent the data can be
consistently explained by
oscillations alone and whether a second sterile neutrino is necessary
to achieve a satisfactory fit \cite{Kopp:2013vaa}, which would not be possible
to accommodate in our setup).
From Fig.~\ref{fig:hdm} we find that the regions allowed by
the HDM signal and neutrino oscillations
indeed overlap, if  only at the $2\sigma$ level.
We note that the corresponding range of $\Delta\Neff$ independently
requires $m_\chi\gtrsim1$\,TeV, as also favored from Fig.~\ref{fig:mv_mchi}.


\section{Discussion}

Before standard neutrino decoupling at $T\sim1$\,MeV, the effective mixing angle $\theta_m$ 
between active and sterile neutrinos is strongly suppressed due to the matter potential generated 
by the  \Uonex\ couplings of the sterile neutrinos \cite{Dasgupta:2013era} 
(see also \cite{Hannestad:2013ana}). 
As the Universe cools, the effective mixing angle eventually reaches its vacuum value~$\theta$.
This may give rise to an additional production of sterile neutrinos due to their \Uonex\ interaction.
The largest effect on the scenario sketched above would result if  the neutrinos completely 
re-thermalized, creating a thermal $N$-$\nu$ bath. 

In that case, conservation of entropy density allows us to determine the temperature
$T_{N\nu}$ of the newly established neutrino bath as $4T_{N\nu}^3=\left[\Neff^\text{SM} +  \left(\Delta\Neff|_\text{bbn}^\text{max}\right)^\frac{3}{4}\right]\left(T_\nu^{\Lambda\text{CDM}}\right)^3$,
where $\Neff^\text{SM} \simeq 3.046$. Rather than Eqns.~(\ref{dNsimp}, \ref{mhdmsimp}), we thus obtain
\begin{align}
\Delta\Neff|_\text{cmb} &=
\frac{1}{4^{\frac13}} \!\left[\Neff^\text{SM} + \left(\Delta\Neff|_\text{bbn}^\text{max}\right)^\frac{3}{4} \right]^\frac{4}{3} - \Neff^\text{SM} \,,\\
m_\text{hdm}^\text{eff} &= \frac{1}{4} \left[\Neff^\text{SM} + \left(\Delta\Neff|_\text{bbn}^\text{max} \right)^\frac{3}{4} \right] m_{N_1}\,.
\end{align}
Rewriting this as $m_{N_1}=2\sqrt{2}\,m_\text{hdm}^\text{eff}/(\Delta\Neff|_\text{cmb}+\Neff^\text{SM})$, 
we immediately see that in the re-thermalization case a sterile neutrino can still consistently 
explain the HDM signal -- but only if its mass is considerably smaller than required by the neutrino 
anomalies.

Turning to potential constraints on our scenario, BBN limits are easily satisfied as already stressed 
earlier. CDM also decouples kinetically too early to imprint  observable dark acoustic oscillation 
(DAO) features in the CMB \cite{Cyr-Racine:2013fsa}. 
Final state $V$ radiation in the decay of SM particles \cite{Laha:2013xua}, finally, does not 
constrain our scenario because  $V$ does not couple to left-handed neutrinos.
An interesting aspect of our HDM component is that it does not necessarily manifest itself as 
perfectly free-streaming particles in the CMB or during structure formation, which in principle can 
be probed~\cite{Archidiacono:2011gq}; by comparing the elastic scattering rate with the Hubble expansion,  
we rather expect complete decoupling only at  
$T_{\nu_R \nu_R}^\text{dpl} \sim 3 \,\text{eV} \fb{m_\chi}{\text{TeV}}{-\frac{2}{3}}  \fb{X_{\nu_R}}{0.2}{-\frac{2}{3}}\fb{m_V^2/X_{\nu_R}}{\text{MeV}^2/0.2}{\frac{2}{3}}$, where the last factor must be of order unity
(see Fig.~\ref{fig:mv_mchi}).

The dominant decay channel of our sterile neutrino is $N_1\rightarrow \nu\nu\nu$. Even though
this is strongly enhanced compared to  the analogous common decay via a virtual $Z$~\cite{Bezrukov:2009th},
we find the resulting lifetime for the best-fit neutrino mixings~\cite{Giunti:2013aea} to be
\be
 \tau_{N_1} \sim 10^5\, t_0  \fb{m_\chi}{\text{TeV}}{-2} \fb{X_{\nu_R}}{0.2}{-2} \fb{m_V^2/X_{\nu_R}}{\text{MeV}^2/0.2}{2} \fb{\text{eV}}{m_{N_1}}{5} 
,
\ee
which greatly exceeds the age of the Universe $t_0$. We note that the decay 
$N_1\rightarrow \nu\gamma$ is even more suppressed due to the necessarily 
small value of $\epsilon$.

\section{Conclusions}

In this article we have considered a mixed DM model as favored by recent cosmological 
observations, which adds a small HDM component to the dominant CDM, the former consisting of 
an eV-scale sterile neutrino and the latter of a TeV-scale Dirac fermion. We have studied the 
cosmological consequences of equipping both these particles with charges under a new 
spontaneously broken \Uonex\ gauge theory, under which all SM particles are singlets.

Thermalizing the \Uonex\ sector in the early universe via the so-called \textit{Higgs portal}  allows 
the thermal production of the CDM\@. The sterile neutrinos would also be thermally produced
and elegantly form the HDM component, essentially because the \Uonex\ sector decouples much 
earlier than 
SM neutrinos.  Remarkably, this is possible for a set of parameters that equip the CDM particles with 
a \Uonex\ mediated self-interaction that is  of the right form and magnitude to provide a 
simultaneous solution to the small-scale problems of $\Lambda$CDM cosmology~\cite{Aarssen:2012fx}. 
Finally, overproduction via mixings is likely prevented by the large thermal potential that the sterile 
neutrinos create by their \Uonex\ interactions~\cite{Hannestad:2013ana,Dasgupta:2013era}; in this case
one can even address the neutrino oscillation anomalies within the same framework.
In other words, a sterile neutrino as preferred by neutrino oscillation anomalies would not only be reconciled with cosmology but promoted to the desired HDM component.

\subsection*{Acknowledgements}
We thank Carlo Giunti, Steen Hannestad, Andreas Ringwald and Neal Weiner for valuable discussions. 
TB would like to thank the German Research Foundation (DFG) for generous support through the Emmy Noether
grant BR 3954/1-1.
JH and JK were supported by the German Research Foundation (DFG) via the
Junior Research Group ``SUSY Phenomenology'' within the Collaborative
Research Center 676 ``Particles, Strings and the Early Universe''.
JH would like to thank the German Academy of Science for support through the Leopoldina Fellowship Programme 
grant LPDS 2012-14.

\medskip
\noindent
{\it Note added.} After the arXiv submission of this work, several 
further concrete models have been proposed that simultaneously address all shortcomings of 
$\Lambda$CDM on small scales \cite{Dasgupta:2013era,Ko:2014bka,Chu:2014lja}, based 
to a varying degree on the  general mechanism suggested in Ref.~\cite{Aarssen:2012fx}.

\bibliography{Interactions}

\end{document}